# Duplicate- and Patient-Overlap Leakage in Brain MRI Tumor Classification: A Multi-Model Evaluation

**Xin Li**[1] · **Samson Oluwadare**[2]

[1] University of Southern California, USA [2] Independent Researcher, Lagos, Nigeria

[1] xli26811@usc.edu [2] sammyoluwadare69@gmail.com

## Abstract

**Background.** Repeated MRI images and distinct slices from the same patient can both compromise evaluation of brain tumor classifiers.

**Objective.** We audited image duplication and examined how patient overlap relates to discrimination, probability error and model ranking.

**Methods.** The Cheng/Figshare dataset contains 3,064 contrast-enhanced T1 MRI slices, three tumor classes and 233 recorded patient IDs. The dataset author advised treating the IDs as individuals, but original DICOM records were unavailable for verification. Four CNN architectures were trained under image-level and recorded-ID-disjoint protocols with five folds matched on size and class counts (40 original runs). A patient-cluster bootstrap quantified uncertainty. The extension added 60 split-seed runs, 40 training-seed runs and 72 fixed-test sibling-exposure runs with class-matched other-patient controls.

**Results.** We found no exact duplicate images and little evidence of cross-patient near-duplicates among screened pairs. Image-level minus recorded-ID-disjoint balanced accuracy ranged from +0.0318 to +0.0462 across models. Expected calibration error and multiclass Brier scores were lower on the image-level evaluation sample. The balanced-accuracy difference persisted across new seeds for ResNet50 and DenseNet121, but their original ranking reversal did not replicate. Only ResNet50 in fold 2 passed Holm correction in the five-test exposure family. All 16 correlated exploratory subgroup estimates were positive, with no consistent dose-response.

**Conclusions.** Patient overlap was associated with higher apparent performance on this benchmark. Fixed-test comparisons could not isolate patient identity from shared anatomy or acquisition characteristics. No independent external clinical validation was performed.

## 1. Introduction

A brain MRI split can put the same image in training and testing. It can also put different slices from the same patient on opposite sides. Even when no pixels are duplicated, those slices can share anatomy, lesion shape and acquisition characteristics. An image-level score may therefore overstate performance on unseen patients (Yagis et al., 2021). Reporting the partitioning unit and test population makes this distinction explicit (Park and Suh, 2024; Tejani et al., 2024).

On the Cheng/Figshare benchmark, the audit found no exact image duplicates. The similarity screen retrieved highly similar slices from the same patient and gave little support for cross-patient near-duplicates among the candidates examined. Even a pair of distinct slices can link training and test data through a shared patient. We therefore compared a protocol that keeps each recorded patient within a single fold with one that allows slices from that patient to cross folds.

Vangala et al. (2026, preprint) traced contamination across redistributed MRI collections and rescored trained models after removing duplicate test images. On the original Cheng/Figshare release, we retrained four CNNs under protocols matched on fold size and class counts, then compared balanced accuracy, calibration, probability error and model rankings. For two architectures, we varied split and training seeds separately. With test slices held fixed, we also compared same-patient sibling donors with same-class donors from other patients.

Figure 1 outlines the primary 40-run comparison. A 172-run extension comprised 100 seed-replication runs and 72 sibling-exposure runs. The exposure and masking analyses are exploratory. Because the audit found no exact duplicates to remove, the study included no deduplication training arm.

### 1.1 Contributions

- We hashed image content, screened candidate near-duplicates and checked patient identifiers. These checks distinguish identical images from distinct, correlated slices and document the limits of the near-duplicate search.
- We compared patient-disjoint and image-level performance for four CNNs, matching the image and class counts in each fold and estimating uncertainty with a paired patient-cluster bootstrap.
- Three additional split allocations and two training seeds tested the stability of protocol differences and model rankings for ResNet50 and DenseNet121.
- With test images held fixed, we compared substitutions at prespecified exposure levels using either same-patient slices or same-class slices from other patients. Regional masking provided an exploratory check on the visual information available to the models.
- We retained the splits, audit results, predictions, reliability bins, bootstrap outputs and scripts that support reconstruction of the primary results. A separate manifest records the extension runs.

## 2. Related Work

### 2.1 Transfer learning for brain-MRI tumor classification

Brain MRI tumor classification studies have evaluated fine-tuned convolutional networks, vision transformers and hybrid models (Rasa et al., 2024; Reddy et al., 2024; Ilani et al., 2025). Interpretation of their scores depends on image reuse, the unit of splitting and the test population. We use four established CNNs to compare evaluation protocols.

### 2.2 Leakage, patient grouping and benchmark comparisons

Different slices from one patient can share visual features useful for classification even when neither image duplicates the other. In brain MRI, slice-wise and subject-wise evaluation can therefore yield different estimates of performance (Yagis et al., 2021). Benchmark design and test composition can also affect the ranking of biomedical imaging methods (Maier-Hein et al., 2018). These findings motivate examining model ordering alongside performance when fold size and class counts are held fixed.

Vangala et al. (2026, preprint) audited duplicate images, patient sharing and acquisition-source confounding across public MRI corpora. In the Nickparvar test split, 95.5% of the images they could trace to Figshare shared a patient with training. That proportion describes a traceable subset of a redistributed collection. The original release used here has recorded IDs for every slice. It still lacks the per-image site information needed to separate patient and acquisition effects.

### 2.3 Duplicate detection and dataset provenance

Image-content hashes identify identical pixel arrays. Thumbnail cross-correlation and perceptual hashing retrieve candidate near-duplicates for full-resolution comparison using normalized cross-correlation, structural similarity and pixel differences. Two neighboring slices from one patient may score highly on these measures while depicting different sections. Conversely, the screen can miss duplicates under cropping, rescaling or other untested transformations. Its findings describe the image pairs examined.

### 2.4 Calibration and uncertainty beyond accuracy

Calibration metrics describe the match between confidence and observed correctness on a specified evaluation population. Under image-level patient overlap, confidence can appear better calibrated on similar images, without establishing useful probabilities for new patients. We report expected calibration error (ECE) and multiclass Brier score alongside discrimination measures. We retained bin-level calibration results and class-specific metrics. The Brier score captures overall probability error and should not be interpreted as a calibration-only statistic.

### 2.5 Methodological positioning

Prior work shows why the unit of splitting matters in brain MRI (Yagis et al., 2021) and why benchmark design can change model rankings (Maier-Hein et al., 2018). Our matched comparisons ask which differences persist when fold size and class counts are held constant. The image audit determines whether the same pixels recur, recorded identifiers establish whether training and test records overlap, and the fixed-test experiment narrows the comparison to the source of training donors. Shared anatomy and acquisition conditions limit any inference about recognition of patient identity.

## 3. Methodology

### 3.1 Study design: two potential leakage pathways

We first assessed duplicate and near-duplicate image contamination as a dataset-integrity question, independent of model scores. We then assessed patient-level dependence using two protocols on the same 3,064 images: A, the original five-fold partition disjoint by recorded patient ID; and B, an image-level partition preserving each fold's image count and class count but permitting patients to appear in training and test. Throughout, patient-disjoint refers to the recorded identifiers. The dataset author advised treating them as distinct individuals, but independent identity verification was unavailable. Each image yielded one pooled out-of-fold (OOF) prediction per model and protocol. The principal contrast is B − A in slice-weighted balanced accuracy. Because different image allocations also change the composition of training and held-out samples, the protocol contrast is not a pure causal effect of patient identity.

Separate replications varied the split or training seed to assess stability. An exploratory intervention held test slices fixed and substituted class-matched training donors to examine sibling exposure. There was no duplicate-removal training arm or synthetic duplication intervention.

### 3.2 Dataset and identity provenance

The public Cheng brain-tumor dataset (Cheng, 2024; Figshare v8, CC BY 4.0) contains 3,064 contrast-enhanced T1 MRI slices from 233 raw patient IDs: 708 meningioma, 1,426 glioma and 930 pituitary-tumor slices. The dataset README cites earlier classification and retrieval studies using this collection (Cheng et al., 2015; Cheng et al., 2016). Individual MATLAB files contain the image, tumor label, PID, tumorBorder and tumorMask. The five-fold indices are distributed separately in cvind.mat, not within each individual .mat record.

The cvind entries were mapped to consecutively numbered files by file order. Under this empirically inferred mapping, none of the 233 recorded IDs crossed folds. The dataset author advised treating the 233 IDs as distinct individuals, consistent with Cheng et al. (2015). The original DICOM records were unavailable for independent verification. The author did not verify the file-order mapping or explain the letter suffixes of MR-prefixed IDs.

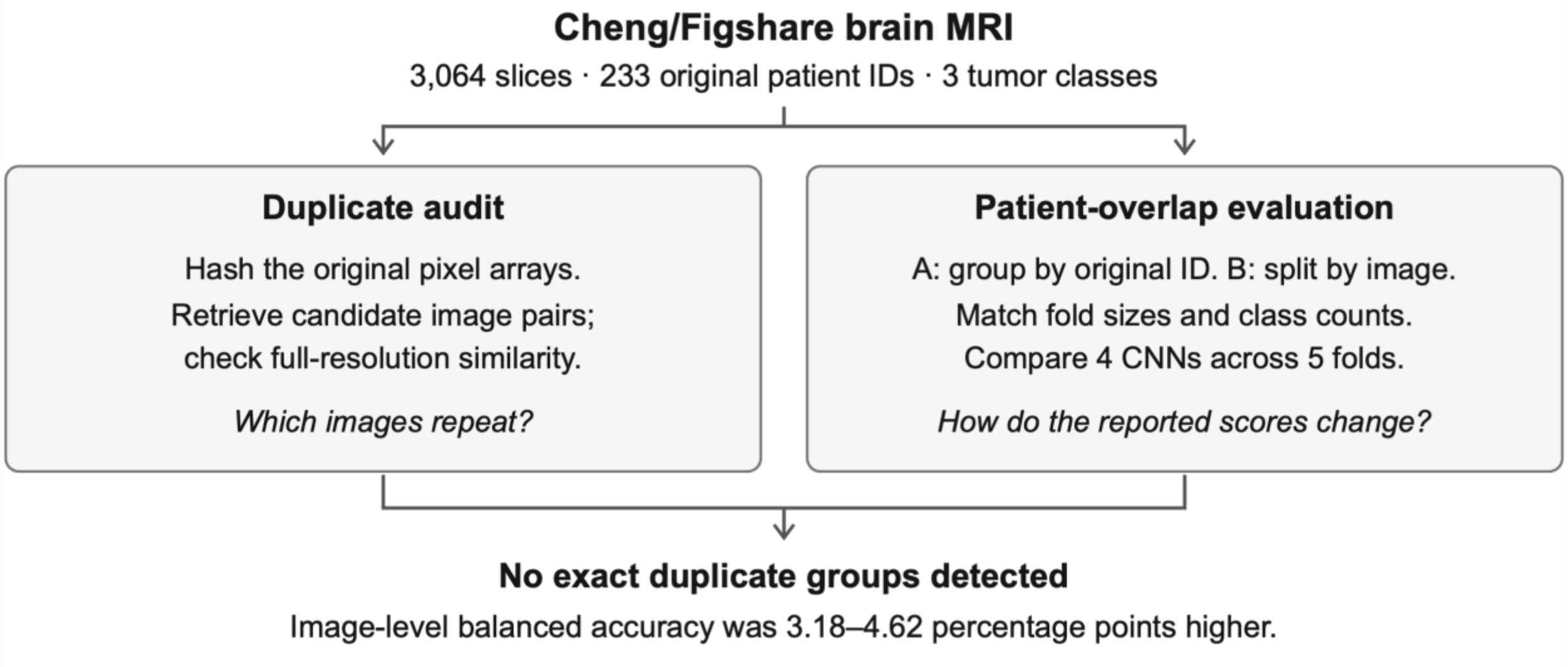

*Figure 1. Two questions on the same 3,064 slices. The image-content audit compared pixel arrays independently of classifier predictions. The four-model evaluation compared matched five-fold protocols using the recorded patient IDs. Figure 6 shows the split-seed comparison and the fixed-test exposure experiment.*

### 3.3 Duplicate-leakage audit: exact hashes and candidate near-duplicates

All 3,064 image records were assessed for duplicate image content. The primary SHA-256 hash covered each array's data type, shape and little-endian pixel bytes in row-major order; it excluded patient identifiers, labels and other metadata. For candidate near-duplicates, 32 × 32 area-resized thumbnails were screened using normalized cross-correlation (NCC) $\geq 0.90$ or 64-bit DCT perceptual-hash Hamming distance $\leq 10$, retrieving 33,432 candidate pairs from approximately 4.69 million image pairs. Candidates were further evaluated by full-resolution NCC, structural similarity (SSIM), normalized pixel-difference measures and equal-pixel fractions. The first-stage thresholds were retrieval rules, not a prospectively fixed label of a near-duplicate.

The audit found zero exact image-content duplicate groups. Among screened cross-patient candidates, the maximum full-resolution NCC was 0.932 and the maximum SSIM was 0.855. No screened cross-patient pair reached NCC $\geq 0.95$ or SSIM $\geq 0.90$. Retrieval recall was unmeasured, including for some image transformations. With no exact groups to remove and no duplicate-removal training experiment, the audit provides no estimate of the performance effect of deduplication.

### 3.4 Patient-overlap identity checks and sensitivity analysis

Removing trailing letters from qualifying MR-prefixed IDs yielded 209 hypothetical stem groups. Eighteen groups merged 42 recorded IDs (811 slices), and 16 groups covering 734 slices spanned the original folds. These stems do not establish that the records belong to the same individual. Using saved predictions, excluding all 734 slices gave a ResNet50 B − A difference of +0.050. All excluded slices were glioma, so exclusion also changed the class mix. Resampling by stem was a second prediction-only sensitivity analysis. No models were retrained on stem-disjoint folds.

### 3.5 Patient-disjoint and image-level evaluation

Protocol A used the original cvind folds. Disjointness was verified at the level of recorded IDs. Each ID occurred in one fold (45, 46, 48, 47 and 47 IDs across folds 1–5). Protocol B assigned images by stratified sampling (seed 11), matching A on slice count and tumor-class counts within each fold (Figure 2). In B, 224 of 233 IDs appeared in multiple folds. Each protocol yielded one out-of-fold prediction per slice and model. The two frozen split CSVs and their SHA-256 hashes define A and B throughout; an earlier design document used the opposite letters. There was no separate validation cohort or validation-based checkpoint selection.

### 3.6 Preprocessing and training

Each raw int16 image was converted to float32, normalized by its own minimum and maximum to [0,1], resized to 224×224 by antialiased bilinear interpolation, replicated into three channels and normalized with ImageNet channel means and standard deviations. Train-only augmentation consisted of horizontal flipping (probability 0.5) and rotation sampled uniformly between −10° and 10° with bilinear interpolation.

We fine-tuned ImageNet-1K V1 pretrained ResNet50, MobileNetV2, EfficientNet-B0 and DenseNet121 backbones after replacing their classification heads with a three-logit linear head. All models used unweighted cross-entropy, batch size 32, 10 epochs, AdamW (learning rate $1\times10^{-4}$, weight decay $1\times10^{-4}$) and a per-iteration cosine schedule. Original training seed 0 was reset for each fold. We evaluated the final epoch for every run, with no early stopping or architecture-specific tuning. Computation used PyTorch on Apple MPS; exact bitwise determinism is not guaranteed.

### 3.7 Primary metrics and calibration

Balanced accuracy (BA) was the mean of class-specific recalls for the three tumor types, calculated from pooled OOF predictions. Macro-F1 was the arithmetic mean of class-wise F1. We also recorded pooled accuracy, class-specific precision/recall/F1, confusion matrices, and prediction disagreement between protocols.

Expected calibration error (ECE) summarizes the gap between accuracy and confidence over 15 equally spaced bins. The multiclass Brier score measures overall squared probability error across the three classes.

$$\mathrm{ECE} = \sum_{b=1}^{15} \frac{n_b}{N} |\mathrm{acc}(b) - \mathrm{conf}(b)|$$

$$\mathrm{Brier} = \frac{1}{N}\sum_{i=1}^{N} \sum_{c=1}^{3} (p_{ic} - 1[y_i = c])^2$$

Here N is the number of evaluated slices, $n_b$ is the count in bin b, and acc(b) and conf(b) are its accuracy and mean confidence. For slice i, $p_{ic}$ is the predicted probability of class c and $y_i$ is the observed class. We retained the bin-level values and the corresponding reliability diagrams. Neither metric measures clinical utility.

### 3.8 Patient-level uncertainty and rankings

For each of 5,000 bootstrap draws (seed 2026), we sampled recorded patient IDs with replacement within tumor class and retained all slices of each sampled ID. We reused the same patient multiplicities across models and protocols, calculated the metric differences within each draw and took the 2.5th and 97.5th percentiles. Thus slices contribute to class recall while IDs are the resampling units. Intervals condition on the fitted models and their predictions. We assessed split and training variability separately. Top-rank frequencies and Kendall's rank correlation are descriptive.

### 3.9 Reproducibility controls

The frozen original evaluation used four models × five folds × two protocols = 40 runs: ten ResNet50 runs and 30 runs across MobileNetV2, EfficientNet-B0 and DenseNet121. We retained split manifests and their SHA-256 hashes, OOF predictions, run metadata, metrics, per-class and reliability tables, bootstrap outputs and verification scripts. Using the saved predictions, we verified 200 numerical and integrity conditions without retraining.

### 3.10 Split- and training-seed replication

For ResNet50 and DenseNet121, three new split seeds (101, 202, 303) generated grouped A folds over the 233 recorded IDs and corresponding margin-matched image-level B folds. With training seed 0 this added 60 runs. The historical A folds came from cvind; seed 11 applies only to historical B. Separately, training seeds 1 and 2 on the original A/B allocations added 40 runs. These replications describe variation on the same cohort. Within-replicate patient bootstrap intervals do not estimate that variation.

### 3.11 Controlled patient-sibling exposure design

A fixed-test exploratory intervention used original patient-disjoint folds 2 and 4, ResNet50 and DenseNet121, and independent training seeds 0 and 1. Eligible held-out patients had at least four slices; for each fold, fixed evaluation slices were removed from the donor pool and were identical across conditions. The CLEAN condition contained no sibling slices from test patients. For exposure levels $L \in \{25\%, 50\%, 75\%, 100\%\}$ of eligible test patients, CTRL added same-class, other-patient donor slices, whereas SIB substituted an equal number of other slices from the selected test patients. Arms were matched for training image count and tumor-class counts, with patient-ID format matched where feasible. The SIB arms intentionally introduced patient overlap and are never labeled leakage-free.

For each model and fold, we averaged the two training seeds' SIB − CTRL differences within each patient bootstrap draw, holding the seeds fixed. We resampled all fixed-test IDs within class and applied eligibility or exposure indicators within each draw. The frozen protocol specified balanced accuracy as the primary metric and five tests for Holm correction: four model–fold contrasts at 100% exposure and a pooled dose-response slope. At full exposure, the four contrasts covered all eligible patients. The protocol did not uniquely specify which contrast to fit for the slope. Our analysis averaged four model–fold slopes of the exposed-subgroup SIB − CTRL differences at 25%, 50%, 75% and 100% exposure. This choice limits confirmatory interpretation of that component. A separate exploratory difference-in-differences slope used the 25% to 75% levels. Two-sided empirical p-values were twice the smaller fraction of draws at or below zero or at or above zero, capped at one. The Holm adjustment applies only to the stated five tests. The smaller reserved-donor training sets preclude direct comparison with historical scores.

### 3.12 Tumor-region test-time diagnostics

For 44 retained final-epoch checkpoints, an exploratory test-time probe used official tumor masks to generate full-image, tumor-only, tumor-plus-margin, non-tumor-only, tumor-removed and shifted-hole artifact-control inputs. Masks were not provided during model training or conventional inference. Because networks were trained on unmasked images, strong masking produces distribution shift; observed responses cannot be equated with causal localization or recognition of patient identity.

## 4. Results

### 4.1 Duplicate-screening findings and patient-level dependence

The audit found zero exact image-content duplicate groups among the 3,064 readable source images. Each of the 233 recorded patient IDs was associated with one tumor class. Among the 33,432 pairs retrieved for detailed comparison, we found highly similar neighboring slices from the same patient. The screened cross-patient pairs provided little evidence of a near-identical duplicate cluster under the assessed criteria. Distinct slices could therefore share patient context even without repeated pixel arrays. Each recorded ID stayed within one original grouped fold, while IDs could cross folds in the matched image-level partition.

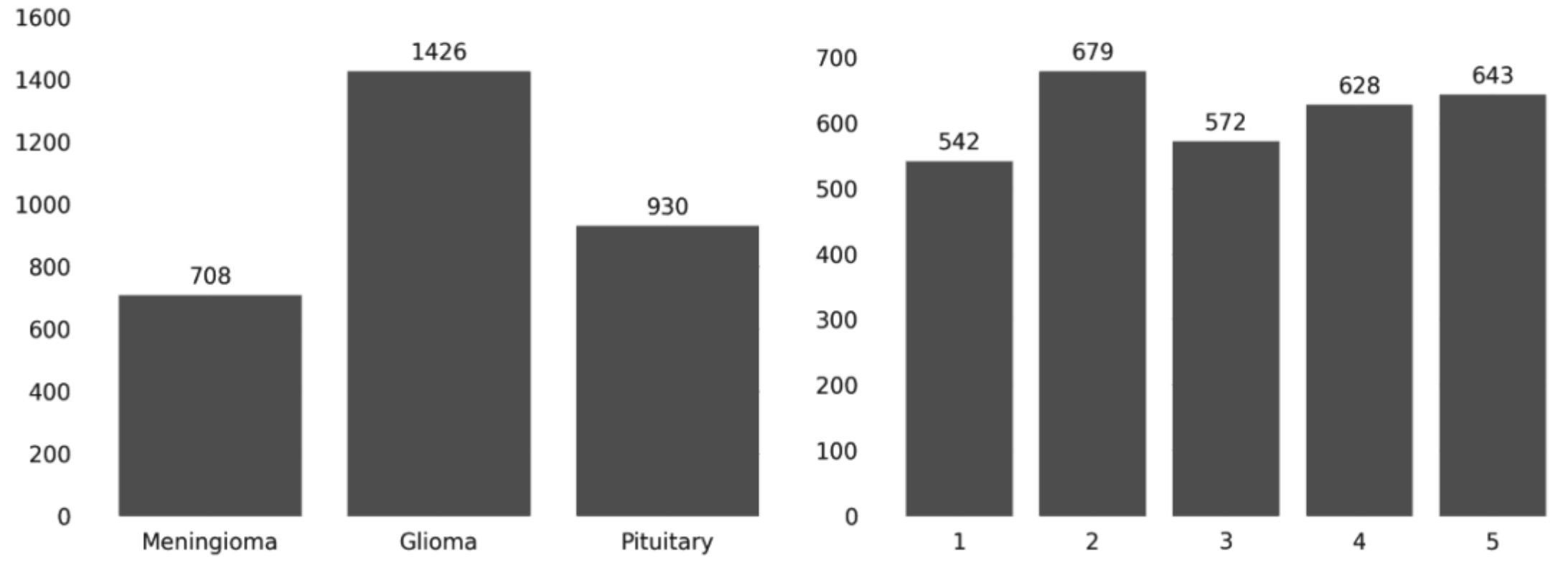

*Figure 2. Slice counts by tumor class (left) and patient-disjoint test fold (right). The 3,064 slices comprise 708 meningioma slices, 1,426 glioma slices and 930 pituitary-tumor slices. Folds 1–5 contain 542, 679, 572, 628 and 643 slices from 45, 46, 48, 47 and 47 distinct patient IDs, respectively.*

### 4.2 Multi-model performance under patient-overlapping evaluation

All four models had higher balanced accuracy under image-level evaluation than under patient-disjoint evaluation at the original seed (Table 1 and Figure 3). The ResNet50 difference was +0.0430 (95% paired patient-cluster bootstrap CI 0.0237–0.0673). The corresponding differences and intervals were +0.0462 (0.0280–0.0685) for MobileNetV2, +0.0420 (0.0238–0.0631) for EfficientNet-B0 and +0.0318 (0.0143–0.0528) for DenseNet121. Section 4.5 examines the stability of the contrast for two of these architectures.

*Table 1. Balanced accuracy in the original 40 runs. A denotes evaluation disjoint by recorded patient ID and B denotes image-level evaluation. Brackets give paired 95% patient-cluster bootstrap confidence intervals for B − A.*

| Architecture | Patient-disjoint BA (A) | Image-level BA (B) | B − A (95% CI) |
|---|---|---|---|
| ResNet50 | 0.9386 | 0.9816 | +0.0430 [0.0237, 0.0673] |
| MobileNetV2 | 0.9246 | 0.9708 | +0.0462 [0.0280, 0.0685] |
| EfficientNet-B0 | 0.9252 | 0.9672 | +0.0420 [0.0238, 0.0631] |
| DenseNet121 | 0.9456 | 0.9774 | +0.0318 [0.0143, 0.0528] |

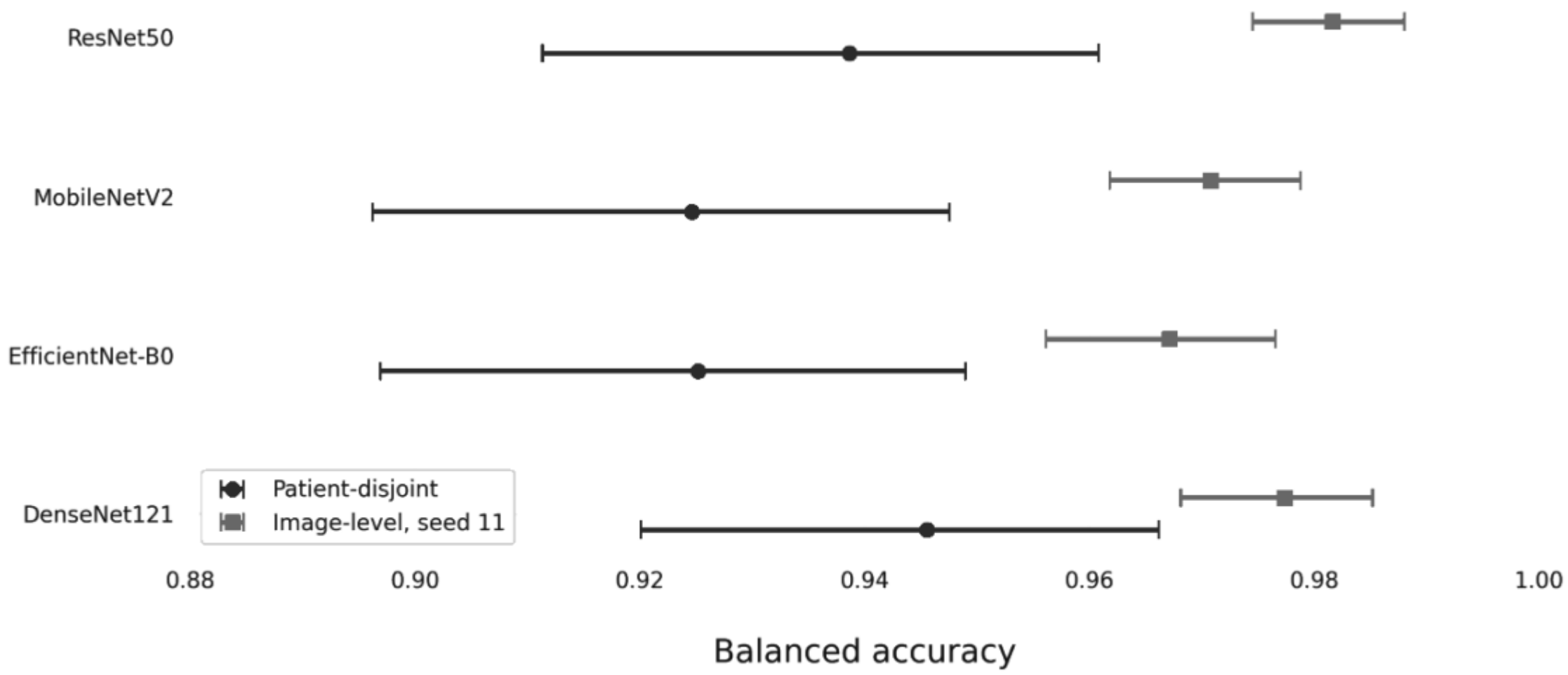

*Figure 3. Balanced accuracy for each architecture under recorded-ID-disjoint evaluation (A, circles) and image-level evaluation (B, squares, seed 11), matched to protocol A in fold size and tumor-class counts. Horizontal bars are within-protocol 95% patient-cluster bootstrap intervals. Table 1 gives paired intervals for the B − A contrast.*

### 4.3 Calibration and probability error

Image-level evaluation also yielded lower ECE and multiclass Brier scores for each of the four original models (Table 2 and Figure 4). For ResNet50, ECE decreased from 0.0321 to 0.0034 and the Brier score from 0.0879 to 0.0224. The paired differences were −0.0287 for ECE (95% CI −0.0456 to −0.0134) and −0.0655 for the Brier score (−0.1007 to −0.0376). These differences describe the evaluated slices. Calibration in an independent clinical cohort was not assessed.

*Table 2. Expected calibration error (ECE) and multiclass Brier score in the original evaluation. Arrows run from patient-disjoint (A) to image-level (B) evaluation. Lower values indicate smaller errors; the Brier score measures overall probability error.*

| Architecture | ECE A → B | Brier A → B |
|---|---|---|
| ResNet50 | 0.0321 → 0.0034 | 0.0879 → 0.0224 |
| MobileNetV2 | 0.0383 → 0.0093 | 0.1047 → 0.0387 |
| EfficientNet-B0 | 0.0320 → 0.0076 | 0.0993 → 0.0379 |
| DenseNet121 | 0.0228 → 0.0040 | 0.0755 → 0.0257 |

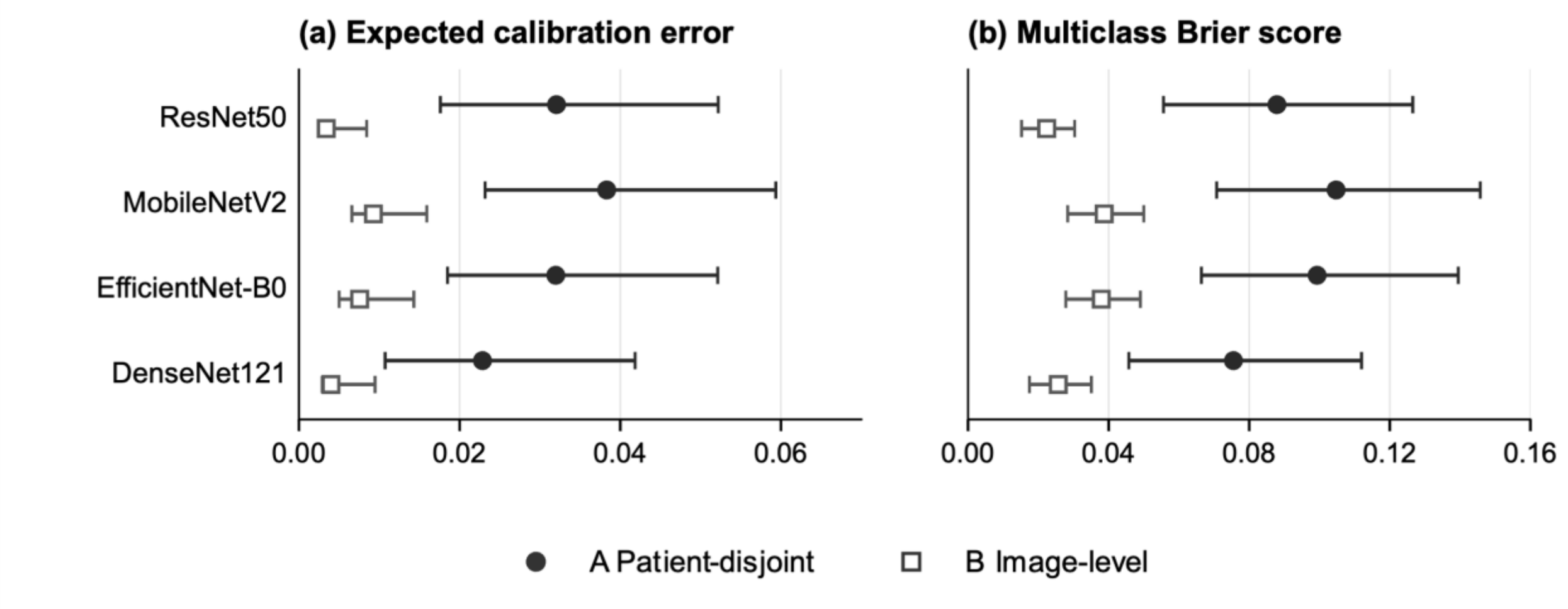

*Figure 4. Expected calibration error (a) and multiclass Brier score (b) for the two evaluation protocols. Points give pooled out-of-fold estimates; horizontal bars give 95% patient-cluster bootstrap confidence intervals. We retained the underlying reliability-bin data.*

## 4.4 Error patterns and model-selection stability

Meningioma recall was higher under image-level evaluation. For ResNet50 it increased from 0.8814 under A to 0.9562 under B. Figure 5 gives the confusion matrices for all four architectures under both protocols. DenseNet121 had the highest observed balanced accuracy under A, and ResNet50 had the highest point estimate under B. For ResNet50 minus DenseNet121, however, the paired confidence intervals included zero under both A (−0.0069, 95% CI −0.0160 to 0.0015) and B (+0.0043, −0.0015 to 0.0104). These contrasts use unrounded estimates.

These intervals leave the relative ordering uncertain. Section 4.5 examines whether the reversal recurs across three new split allocations or either new training seed.

## 4.5 Split- and training-seed robustness of the patient-overlap contrast

Across the new split and training seeds, image-level minus patient-disjoint balanced accuracy ranged from +0.0306 to +0.0471 for ResNet50 and DenseNet121. Figure 6a shows the split-seed estimates. The historical ranking reversal did not recur in any of the three new split allocations or either new training seed.

DenseNet121 had the higher balanced-accuracy point estimate under both protocols at each of the three new split seeds. On the original splits, the image-level ResNet50 minus DenseNet121 difference was +0.0043 at training seed 0, −0.0039 at seed 1 and −0.0042 at seed 2. All three paired intervals included zero. Across historical and new seeds, sample standard deviations of balanced accuracy were approximately 0.001–0.005 for split seeds and 0.002–0.004 for training seeds.

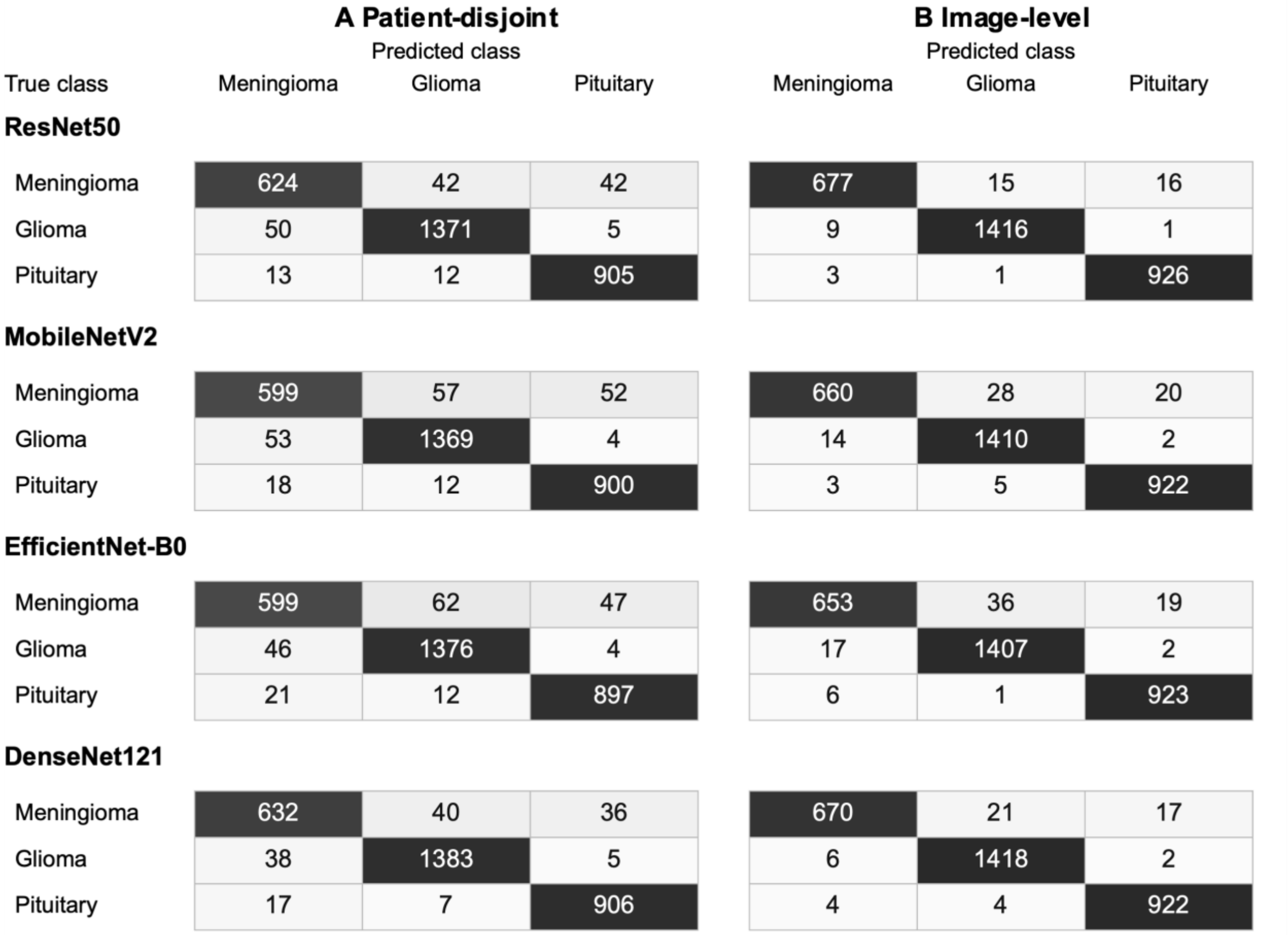

*Figure 5. Out-of-fold confusion matrices. Each model is shown under patient-disjoint (A) and image-level (B) evaluation. Rows give the true tumor class and columns the predicted class. Numbers are slice counts; shading shows the proportion within each true class on the same scale across all eight matrices.*

### 4.6 Controlled sibling exposure and regional diagnostics

The 72 fixed-test runs spanned two models, two folds, two training seeds and nine conditions. At 100% exposure (Figure 6b), the SIB − CTRL balanced-accuracy differences for ResNet50 were +0.059 (95% CI 0.026 to 0.104) in fold 2 and +0.002 (−0.013 to 0.015) in fold 4. For DenseNet121 they were +0.045 (0.007 to 0.093) and +0.019 (−0.002 to 0.044), respectively. Only ResNet50 in fold 2 passed Holm correction at 0.05. None of its 5,000 bootstrap draws was at or below zero, which limits the precision of the tail estimate. The adjusted p-value for DenseNet121 in fold 2 was 0.072. Using the implemented sibling contrast, the pooled slope per 100 percentage points was −0.010 (95% CI −0.073 to 0.031, adjusted p = 1.00). This estimate provided no evidence of a consistent dose-response under the implemented definition.

Across nested exposure levels, all 16 exploratory SIB − CTRL point estimates in the exposed subgroup were positive. These comparisons are dependent and fall outside the five-test Holm family. Seven nominal 95% intervals excluded zero when all fixed-test IDs were resampled and each draw was restricted to the exposed subgroup. Eight excluded zero when only exposed IDs were resampled. Exploratory spillover contrasts showed no consistent pattern. The comparator used here included ineligible patients. At 100% exposure, the 3–6 patients per fold who remained unexposed were ineligible for sibling exposure.

Across 44 checkpoints, balanced accuracy was approximately 0.80–0.93 on non-tumor-only inputs and 0.36–0.48 on tumor-only masked inputs. The tumor-only inputs differed substantially from the unmasked images used for training, limiting what the scores can tell us about localization. Some sibling-exposure differences persisted with non-tumor context. They could reflect patient identity, scanner or site characteristics, tumor location or other correlated factors. These probes cannot distinguish among those explanations.

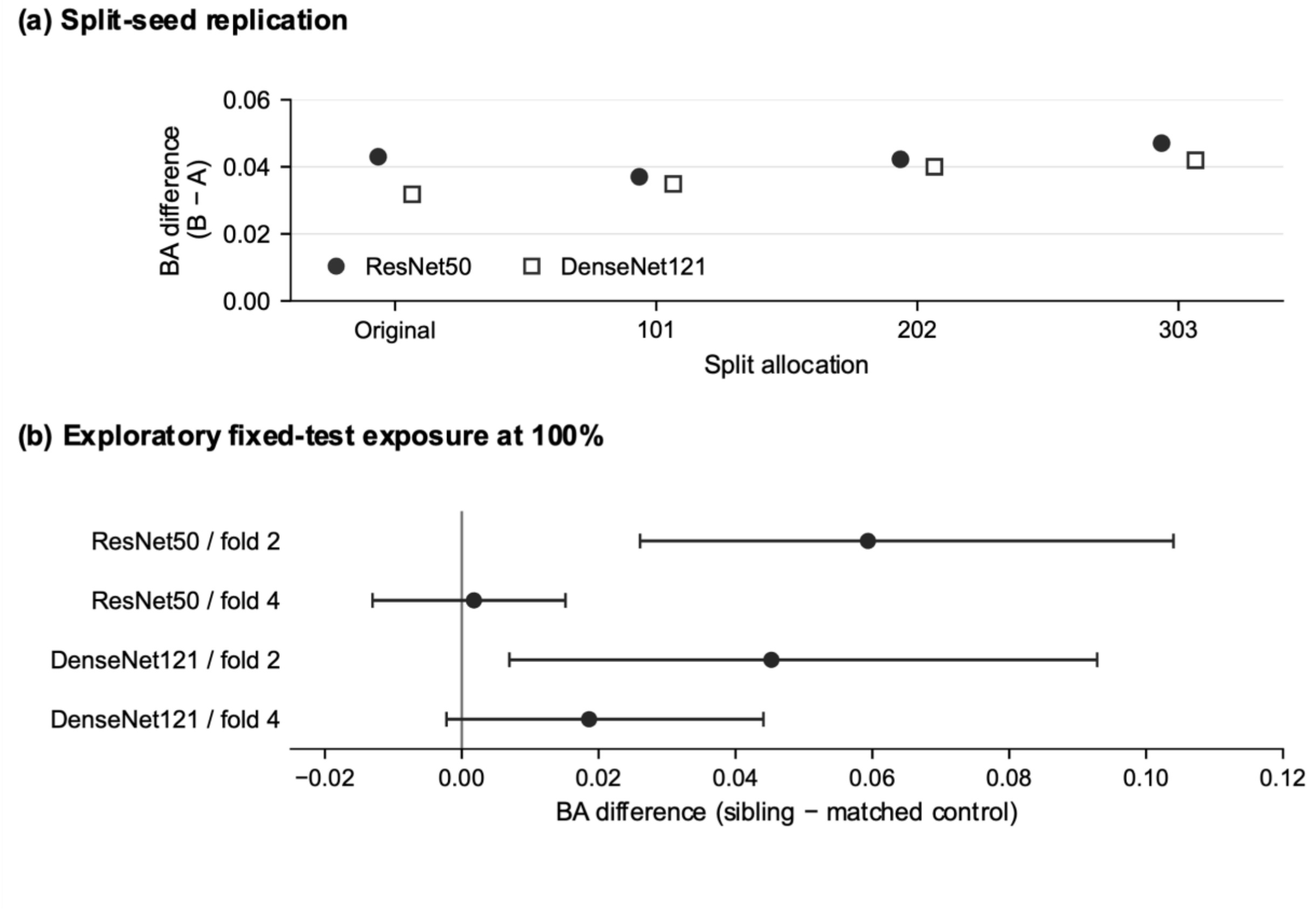

*Figure 6. (a) Descriptive split-seed replication. Points show image-level minus patient-disjoint balanced accuracy at each split allocation. The original comparison uses cvind folds for A and seed 11 for B. (b) Exploratory fixed-test exposure study. Points compare same-patient siblings with matched other-patient controls at 100% exposure. Bars are nominal 95% patient-cluster bootstrap intervals. Of four model–fold contrasts and the pooled slope, only ResNet50 in fold 2 passed Holm correction at 0.05.*

### 4.7 Evidence synthesis across leakage pathways

The matched protocol comparison yielded a consistent difference across four models and across additional seeds for the two models retested. The fixed-test comparison narrowed the question to sibling exposure, for which estimates varied by fold and showed no consistent dose-response. Neither comparison estimates what removing duplicate images would have done.

# 5. Discussion

## 5.1 Principal findings: distinguishing duplicate and patient overlap

An MRI benchmark can contain distinct pixel arrays yet evaluate a model on patients whose other slices it has seen. On this cohort, allowing patient overlap was associated with approximately 3–5 percentage points higher balanced accuracy across the four original architectures. The difference persisted across new seeds for the two architectures retested, even as their relative ranking changed. An exact-duplicate audit alone would miss this distinction, and the near-duplicate screen found little evidence of cross-patient near-identical images among the candidates it retrieved.

## 5.2 What matching fold size and class counts controls

Matching fold size and tumor-class counts ruled out differences in those two margins as explanations for the protocol contrast. Both protocols used the same training settings, and pairing by recorded patient ID accounted for the shared cohort when estimating uncertainty. The particular slices assigned to training and testing still differed. Holding the test slices fixed in the donor experiment removed this source of variation on the test side while allowing the source of the training slices to change.

This distinction also matters when comparing our estimates with those of Vangala et al. (2026, preprint), who rescored trained models after removing duplicates from a test set. Here, models were retrained under each protocol and the same 3,064 slices received out-of-fold predictions under both. Pairing those predictions by original ID quantifies a difference between evaluation protocols on this cohort. It cannot attribute that difference solely to patient identity.

## 5.3 Ranking and calibration implications

The within-model difference between protocols persisted across seeds for the two leading architectures. The between-model difference under image-level evaluation was small enough to change sign, making a ranking from one image-level split a poor basis for choosing between them. Lower ECE describes closer agreement between confidence and observed accuracy across the specified bins. A lower multiclass Brier score indicates less overall probability error. Both describe the evaluated population. Neither establishes calibration on an independent clinical cohort.

## 5.4 Interpreting patient-sibling exposure without conflating duplicate slices

Every exposure condition used the same held-out slices, so differences in test-set composition cannot explain a SIB − CTRL contrast. The training substitution nevertheless changes more than access to a recorded patient. Sibling slices can share acquisition characteristics, anatomy and lesion location. All 16 exploratory estimates were positive across nested conditions, but their magnitudes varied by fold, only one prespecified contrast passed Holm correction, and there was no consistent dose-response. The regional probes could not distinguish patient identity from these other shared cues.

## 5.5 Relation to duplicate and patient-leakage literature

The distinction between slice-wise and subject-wise evaluation in brain MRI (Yagis et al., 2021), the sensitivity of biomedical rankings to benchmark design (Maier-Hein et al., 2018), and evidence of multiple contamination pathways in public brain MRI collections (Vangala et al., 2026, preprint) have a common implication for reporting. A reader needs to know both how images were checked for reuse and whether test patients could contribute other slices to training. The matched comparisons on Cheng/Figshare illustrate why the answer to one question cannot stand in for the other.

### 5.6 Limitations of the duplicate and patient-overlap evidence

The patient-disjoint designation applies to the 233 original IDs that the dataset author advised treating as individuals. The absence of original DICOM records leaves independence across differently suffixed IDs unverified. The 209 hypothetical stem groups therefore provide a sensitivity analysis only. MR-prefixed IDs in this release occur only in glioma. The file-order mapping to folds was inferred, and per-image acquisition-site labels were unavailable. Exact-duplicate findings apply to the pixel arrays examined. The two-stage near-duplicate screen has no validated sensitivity for all transformations or universal diagnostic threshold.

All four models were compared on one original pair of allocations with a single training seed. Only ResNet50 and DenseNet121 underwent replication, so the stability of the protocol contrast across seeds is specific to these two architectures and this cohort. The sibling-exposure study covered two folds and two training seeds, with just one prespecified model–fold contrast passing multiplicity correction. The donors could share anatomy, lesions or acquisition context with test images, and masking introduced severe distribution shift. Patient-cluster bootstrap intervals describe uncertainty conditional on the fitted predictions in this cohort. Generalization to an independent hospital or clinical population was not evaluated.

Testing the near-duplicate screen on known transformations would clarify its retrieval limits. An independent, patient-identified cohort would be needed to assess transfer to another clinical setting.

## 6. Conclusion

For claims about unseen patients, brain MRI benchmarks should document image-duplication checks, separate recorded patients across training and testing, and report uncertainty alongside discrimination and calibration. On Cheng/Figshare, image-level evaluation was associated with higher apparent balanced accuracy and lower probability error even though the audit found no exact duplicates and little evidence of cross-patient near-duplicates among screened pairs. The balanced-accuracy contrast persisted across additional seeds for two models. Their original ranking reversal did not recur, and the fixed-test experiment left the mechanism unresolved.

## Data and Code Availability

The Cheng brain tumor dataset (Figshare v8, CC BY 4.0) is publicly available at https://doi.org/10.6084/m9.figshare.1512427.v8. The analysis code, derived split manifests, audit records, out-of-fold predictions, bootstrap outputs, and verification scripts are not included in this preprint release. Large trained model checkpoints are also not included.

## Declaration of Generative AI Use

Generative AI tools assisted with programming, manuscript drafting and editing, document preparation and verification. Codex and Claude Code assisted with editing and checking versions of this preprint. The authors take responsibility for all scientific content, analyses and the final text.